\documentclass[aps,prl,reprint,superscriptaddress,nofootinbib,longbibliography,nobalancelastpage]{revtex4-2}
\usepackage[T1]{fontenc}
\usepackage{lmodern}
\usepackage{amsmath,amssymb,bm,booktabs,xcolor}
\usepackage{microtype}

\usepackage{graphicx}
\usepackage{tikz}
\usetikzlibrary{decorations.pathmorphing}
\usepackage[colorlinks=true,linkcolor=blue!50!black,citecolor=blue!50!black,urlcolor=blue!50!black]{hyperref}
\usepackage{orcidlink}

\newcommand{\dd}{\mathrm{d}}
\newcommand{\ii}{\mathrm{i}}
\newcommand{\Id}{\mathbf{1}}

\begin{document}
\title{Nonlinear Response of Schwarzschild Black Holes at Arbitrary Perturbative Order}
\author{Jiaqi Chen\,\orcidlink{0000-0003-3537-3846}}
\email{jiaqichen@cup.edu.cn}
\affiliation{Beijing Key Laboratory of Optical Detection Technology for Oil and Gas, China University of Petroleum--Beijing, Beijing 102249, China}
\affiliation{Basic Research Center for Energy Interdisciplinary, College of Science, China University of Petroleum--Beijing, Beijing 102249, China}
\author{Yu-Cun Xie\,\orcidlink{0000-0002-7603-7411}}
\email{yzx5296@psu.edu}
\affiliation{Institute for Gravitation and the Cosmos, The Pennsylvania State University, University Park, Pennsylvania 16802, USA}
\affiliation{Department of Physics, The Pennsylvania State University, University Park, Pennsylvania 16802, USA}
\date{October 20, 2026}
\begin{abstract}
We establish recursive integration-by-parts identities that symbolically reduce the inhomogeneous solutions of the Regge-Wheeler-Zerilli equation to a small set of master integrals, yielding the nonlinear response kernels of Schwarzschild black holes at arbitrary perturbative order. We further derive a closed ODE system that enables rapid, accurate, and stable numerical evaluation of these master integrals. To demonstrate our result, we compute quadratic, cubic, and quartic quasinormal-mode coupling coefficients as special cases of the response kernels.
\end{abstract}
\maketitle

Gravitational waves probe the nonlinear dynamics of black holes~\cite{Berti:2009kk}. Nonlinear wave couplings generate radiation at combinations of the driving frequencies~\cite{Zlochower:2003yh}. A response kernel gives the strength and phase of each coupling: multiplying it by the product of the parent amplitudes yields the generated complex amplitude. The same description applies to nonlinear wave scattering and to couplings between quasinormal modes (QNMs), the black hole's damped natural oscillations.

Second-order scattering responses have been calculated for Schwarzschild black holes~\cite{Cardoso:2026llh,Bucciotti:2026jhr} and measured in simulations of Kerr black holes~\cite{Keeble:2026epf}. Nonlinear ringdown has been identified in black-hole merger simulations~\cite{Cheung:2022rbm,Mitman:2022qdl,Khera:2023horizon,Dyer:2026content}, and perturbative calculations have yielded quadratic QNM amplitudes~\cite{Perrone:2024analytic,RedondoYuste:2024spin,Ma:2024qnm,Bucciotti:2024jrv,Bourg:2024jme,Khera:2025qnm,Bourg:2025lpd} and, more recently, cubic amplitudes for Kerr black holes~\cite{Khera:2026lna}.

At higher orders, waves generated by earlier interactions themselves become sources of new radiation. A second-order wave can combine with a first-order wave to produce a third-order signal, alongside the direct interaction of three first-order waves. For Schwarzschild black holes, the radiative equations contain the same Regge--Wheeler--Zerilli (RWZ) operator at every order, with nonlinear interactions entering through sources built from lower-order fields~\cite{Regge:1957td,Zerilli:1970se,Brizuela:2006ne,Brizuela:2007zza,Brizuela:2009qd,Spiers:2023mor}. Solving these equations introduces integrals over the sources. The sources grow rapidly in size and become numerically delicate, while the unregularized Green-function integrals need not converge.

In quantum field theory, integration-by-parts (IBP) reduction is a standard technique for expressing rational integrals in terms of a finite master basis~\cite{Tkachov:1981wb,Chetyrkin:1981qh}. This is typically combined with the differential-equation method, which supplies rapid numerical evaluation of the master integrals~\cite{Kotikov:1990kg,Remiddi:1997ny}. Chen and Feng first generalized this framework to curved spacetime for de Sitter correlators with Hankel-function integrands~\cite{Chen:2023iix}. Subsequent work improves the boundary calculation of differential equations~\cite{Chen:2024glu}.

Inspired by this framework, we establish an IBP reduction for the sourced RWZ hierarchy, deriving recursive IBP identities that reduce nonlinear response integrals at every perturbative order to a finite basis of master integrals. When products of radial fields are differentiated, the field equations eliminate second derivatives and, for driven intermediate fields, introduce their sources. Repeated source substitutions preserve the total perturbative order and terminate once all field factors in the sources are first order. Applying the same identities to derivatives with respect to an auxiliary parameter then yields a closed system of differential equations for the master integrals.

We carry out the source construction, boundary regularization, and integral reduction analytically, retaining symbolic angular labels and frequencies until numerical evaluation. The construction treats arbitrary complex frequencies, including purely damped couplings and exact resonant limits. The resulting symbolic formulas can be reused for rapid evaluation of different responses. As an application, we specialize the parent waves to QNMs and compute quadratic, cubic, and quartic coupling coefficients, including direct and cascaded contributions.

We use units $G=c=2M=1$, areal radius $r$, and time dependence $e^{-\ii\omega t}$.
Here $M$ is the black-hole mass, and $(\ell,m)$ label spherical harmonics.
The horizon lies at $r=1$. With $f=(r-1)/r$,
$D_*=f\partial_r$ differentiates with respect to the tortoise coordinate $r_*=r+\ln(r-1)$.

\emph{Tensor perturbation hierarchy.---}
Let $g_0$ be the Schwarzschild metric, satisfying $G[g_0]=0$, and expand
\begin{equation}
 g(\epsilon)=g_0+\sum_{N\ge1}\epsilon^Nh^{(N)} .
 \label{eq:metricexpansion}
\end{equation}
Let $G^{(n)}$ denote the symmetric $n$-linear part of the Einstein tensor expanded about $g_0$, with $1/n!$ included in its definition~\cite{Spiers:2023mor}. Then
\begin{equation}
 \begin{split}
 G[g]={}&G[g_0]+\epsilon G^{(1)}[h^{(1)}]\\
 &+\epsilon^2\bigl(G^{(1)}[h^{(2)}]
 +G^{(2)}[h^{(1)},h^{(1)}]\bigr)+O(\epsilon^3).
 \end{split}
 \label{eq:einsteinexpansion}
\end{equation}
Setting each coefficient to zero gives
\begin{equation}
 G^{(1)}[h^{(1)}]=0,\qquad
 G^{(1)}[h^{(2)}]=-G^{(2)}[h^{(1)},h^{(1)}].
 \label{eq:ordersolution}
\end{equation}
At order $N\ge2$, the unknown field $h^{(N)}$ satisfies
\begin{equation}
 G^{(1)}[h^{(N)}]
 =\mathcal T^{(N)}[h^{(1)},\ldots,h^{(N-1)}].
 \label{eq:einstein}
\end{equation}
The source $\mathcal T^{(N)}$ contains only known lower-order fields; gauge and boundary conditions specify the solution for $h^{(N)}$. At third order,
\begin{equation}
 \mathcal T^{(3)}=-2G^{(2)}[h^{(1)},h^{(2)}]
 -G^{(3)}[h^{(1)},h^{(1)},h^{(1)}].
 \label{eq:highervertices}
\end{equation}
These are the $2+1$ cascade and $1+1+1$ direct contributions. At arbitrary order,
\begin{equation}
 \mathcal T^{(N)}=-\sum_{q=2}^{N}
 \sum_{\substack{p_1+\cdots+p_q=N\\p_i\ge1}}
 G^{(q)}[h^{(p_1)},\ldots,h^{(p_q)}].
 \label{eq:generaltensorsource}
\end{equation}
\emph{Master fields and source completion.---}
For $\ell\ge2$, harmonic projection and the constraint equations reduce $G^{(1)}[h^{(N)}]=\mathcal T^{(N)}$ to one RWZ equation in each parity sector, $P=E$ (even) or $O$ (odd).
The same operator governs every order:
\begin{equation}
 \mathcal L_\ell^P\Psi_{\ell m}^{(N),P}=S_{\ell m}^{(N)},\qquad
 \mathcal L_\ell^P=D_*^2+\omega^2-V_\ell^P .
 \label{eq:rwz}
\end{equation}
Here $V_\ell^E$ and $V_\ell^O$ are the Zerilli and Regge--Wheeler potentials, respectively, and $S_{\ell m}^{(N)}$ is the source for the selected daughter parity.
We use the Zerilli--Moncrief (even) and Cunningham--Price--Moncrief (odd) variables as the raw master fields in Regge--Wheeler gauge~\cite{Martel:2005ir,Spiers:2023mor}.
For specified parent modes, $m=\sum_i m_i$ and $\omega=\sum_i\omega_i$, while parity conservation fixes $P$ once the daughter multipole $\ell$ is chosen.
We suppress the daughter labels in the following redefinition.

Near the horizon and infinity, the master field contains terms determined locally by the nonlinear source. We subtract their leading contributions to isolate the free-wave amplitude. Extending the second-order regularization~\cite{Gleiser:1998rw,Nakano:2007cj,Brizuela:2009qd,Bucciotti:2024zyp,Bucciotti:2026jhr}, at each order $N\ge2$ we define
\begin{equation}
 \Psi^{(N)}=\Psi_{\rm raw}^{(N)}-\Delta\Psi^{(N)}.
 \label{eq:masterredefinition}
\end{equation}
Substitution into $\mathcal L\Psi_{\rm raw}^{(N)}=S_{\rm raw}^{(N)}$ fixes the source correction:
\begin{equation}
 S^{(N)}=S_{\rm raw}^{(N)}-\mathcal L\Delta\Psi^{(N)}.
 \label{eq:completion}
\end{equation}
The subtraction $\Delta\Psi^{(N)}$ is a sum of products of lower-order fields with total order $N$.
We fix the coefficients of $\Delta\Psi^{(N)}$ by matching $\mathcal L\Delta\Psi^{(N)}$ to the leading terms of $S_{\rm raw}^{(N)}$ near the horizon and infinity.
Restoring $\Psi_{\rm raw}^{(N)}=\Psi^{(N)}+\Delta\Psi^{(N)}$ before metric reconstruction supplies the perturbation $h^{(N)}$ needed at the next order.

\emph{One radial equation for both parities.---}
For $\ell\ge2$, we combine the Darboux transformation~\cite{Glampedakis:2017rar} with a radial rescaling to introduce our common variable $u^{(N)}$:
\begin{equation}
 u^{(N)}=r^{-3}(r-1)^{\ii\omega}
 \begin{cases}
 \Psi^{(N)},&P=O,\\
 (-D_*+\mathcal D)\Psi^{(N)}/\kappa,&P=E
 \end{cases}.
 \label{eq:ustrip}
\end{equation}
Here $\lambda=(\ell-1)(\ell+2)/2$ and $Z=2\lambda r+3$; the Darboux function $\mathcal D(r)$ and normalization $\kappa$ are given in the End Matter.
Both parities obey the same sourced radial equation,
\begin{equation}
 \begin{gathered}
 (u^{(N)})''+\alpha(u^{(N)})'+\beta u^{(N)}=\widetilde S^{(N)},\\
 \begin{aligned}
 \alpha&=\frac5r+\frac{1-2\ii\omega}{r-1},\\
 \beta&=\omega^2+\frac{-4+5\ii\omega+2\lambda}{r}\\
 &\quad+\frac{4-5\ii\omega+2\omega^2-2\lambda}{r-1} .
 \end{aligned}
 \end{gathered}
 \label{eq:unified}
\end{equation}
Primes denote $r$-derivatives. The coefficients have only simple poles at $r=0,1$.

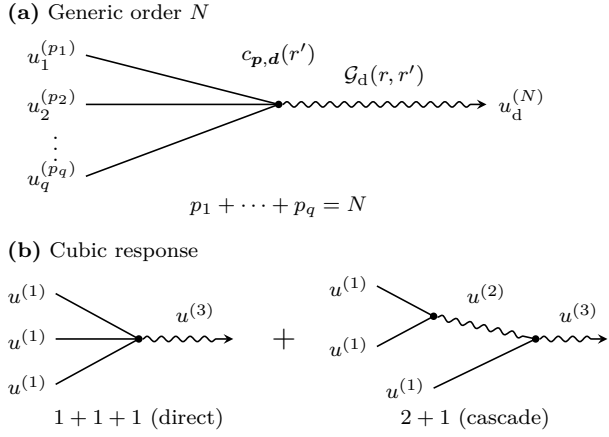
\begin{figure}[tb]
\centering
\begin{tikzpicture}[x=1cm,y=1cm,>=stealth,
 every node/.style={font=\footnotesize,inner sep=1.5pt},
 field/.style={line width=0.6pt},
 prop/.style={line width=0.6pt,decorate,
 decoration={snake,amplitude=0.35mm,segment length=1.9mm,post length=1.5mm}}]
 \node[anchor=west] at (0,4.1) {\textbf{(a)} Generic order $N$};
 \draw[field] (1.1,3.55) -- (3.65,2.9);
 \draw[field] (1.1,2.9) -- (3.65,2.9);
 \draw[field] (1.1,1.95) -- (3.65,2.9);
 \node[anchor=east] at (1.05,3.55) {$u_1^{(p_1)}$};
 \node[anchor=east] at (1.05,2.9) {$u_2^{(p_2)}$};
 \node at (0.7,2.42) {$\vdots$};
 \node[anchor=east] at (1.05,1.95) {$u_q^{(p_q)}$};
 \fill (3.65,2.9) circle (1.4pt);
 \node at (3.65,3.55) {$c_{\bm p,\bm d}(r')$};
 \draw[prop,->] (3.7,2.9) -- (6.4,2.9);
 \node at (5.05,3.28) {$\mathcal G_{\rm d}(r,r')$};
 \node[anchor=west] at (6.5,2.9) {$u_{\rm d}^{(N)}$};
 \node at (3.65,1.55) {$p_1+\cdots+p_q=N$};

 \node[anchor=west] at (0,0.95) {\textbf{(b)} Cubic response};
 \draw[field] (0.7,0.4) -- (1.8,-0.2);
 \draw[field] (0.7,-0.2) -- (1.8,-0.2);
 \draw[field] (0.7,-0.8) -- (1.8,-0.2);
 \node[anchor=east] at (0.65,0.4) {$u^{(1)}$};
 \node[anchor=east] at (0.65,-0.2) {$u^{(1)}$};
 \node[anchor=east] at (0.65,-0.8) {$u^{(1)}$};
 \fill (1.8,-0.2) circle (1.4pt);
 \draw[prop,->] (1.85,-0.2) -- (3.05,-0.2);
 \node at (2.55,0.15) {$u^{(3)}$};
 \node at (1.8,-1.25) {$1+1+1$ (direct)};
 \node[font=\large] at (3.72,-0.2) {$+$};

 \draw[field] (4.95,0.5) -- (5.7,0.1);
 \draw[field] (4.95,-0.3) -- (5.7,0.1);
 \node[anchor=east] at (4.9,0.5) {$u^{(1)}$};
 \node[anchor=east] at (4.9,-0.3) {$u^{(1)}$};
 \fill (5.7,0.1) circle (1.4pt);
 \draw[prop] (5.75,0.09) -- (7.05,-0.2);
 \node at (6.4,0.4) {$u^{(2)}$};
 \draw[field] (5.7,-0.85) -- (7.05,-0.2);
 \node[anchor=east] at (5.65,-0.85) {$u^{(1)}$};
 \fill (7.05,-0.2) circle (1.4pt);
 \draw[prop,->] (7.1,-0.2) -- (8,-0.2);
 \node at (7.62,0.15) {$u^{(3)}$};
 \node at (6.25,-1.25) {$2+1$ (cascade)};
\end{tikzpicture}
\caption{Tree structure of the source in Eq.~\eqref{eq:genericusource}. (a) An order-$N$ vertex joins $q$ lower-order fields, each carrying zero or one radial derivative. Wavy lines denote Green-function propagation. (b) Direct and cascade contributions to the cubic source.}
\label{fig:source-trees}
\end{figure}

\emph{Source structure at arbitrary order.---}
Eliminating higher radial derivatives with the field equations gives the following products of $q$ lower-order radiative fields, illustrated in Fig.~\ref{fig:source-trees}:
\begin{equation}
 \begin{split}
 \widetilde S^{(N)}(r)={}&\sum_{q=2}^{N}
 \sum_{\substack{p_1+\cdots+p_q=N\\p_i\ge1}}
 \sum_{d_1,\ldots,d_q=0}^{1}\\
 &c_{\bm p,\bm d}(r)
 \prod_{i=1}^{q}\bigl(u_i^{(p_i)}\bigr)^{[d_i]}.
 \end{split}
 \label{eq:genericusource}
\end{equation}
Here $\bm p=(p_1,\ldots,p_q)$ records perturbative orders and $\bm d=(d_1,\ldots,d_q)$ records derivatives, with $u^{[0]}=u$ and $u^{[1]}=\partial_r u$.
Sums over allowed harmonic and parity assignments are understood.
Frequency conservation cancels the fractional powers introduced by Eq.~\eqref{eq:ustrip}, leaving $c_{\bm p,\bm d}(r)$ rational in $r$.

\emph{Green-function integral.---}
The subscript $\mathrm d$ labels the daughter channel.
The Green function is built from two homogeneous solutions of Eq.~\eqref{eq:unified}: $u_{\rm d}^H$, ingoing at the horizon, and $u_{\rm d}^{\infty}$, outgoing at infinity~\cite{Leaver:1986response,Lagos:2023generation}.
We obtain them through Eq.~\eqref{eq:ustrip} from physical master solutions with unit wave amplitude at their respective boundaries.
Abel's identity makes the weighted Wronskian
\begin{equation}
 \begin{split}
 \mathcal W_{\rm d}={}&r^5(r-1)^{1-2\ii\omega_{\rm d}}\\
 &\times\bigl[u_{\rm d}^H(u_{\rm d}^\infty)'
 -(u_{\rm d}^H)'u_{\rm d}^\infty\bigr]
 \end{split}
 \label{eq:uwh}
\end{equation}
constant in $r$.
For $\mathcal W_{\rm d}\ne0$, the Green function is
\begin{equation}
 \mathcal G_{\rm d}(r,r')=
 \frac{r'^5(r'-1)^{1-2\ii\omega_{\rm d}}}{\mathcal W_{\rm d}}
 u_{\rm d}^H(r_<)u_{\rm d}^\infty(r_>),
 \label{eq:radialgreen}
\end{equation}
where $r_<=\min(r,r')$ and $r_>=\max(r,r')$.
The driven solution is
\begin{equation}
 u_{\rm d}^{(N)}(r)=\int_1^\infty\dd r'\,
 \mathcal G_{\rm d}(r,r')\widetilde S^{(N)}(r').
 \label{eq:greensolution}
\end{equation}
The outgoing amplitude $B_{\rm d}^{(N)}$ of the completed master field is the coefficient of $u_{\rm d}^{\infty}$ as $r\to\infty$~\cite{Bucciotti:2026jhr}:
\begin{equation}
 \begin{split}
 B_{\rm d}^{(N)}&=\frac{J_N}{\mathcal W_{\rm d}}\\
 &=\frac1{\mathcal W_{\rm d}}\int_1^\infty\dd r\,
 r^5(r-1)^{1-2\ii\omega_{\rm d}}u_{\rm d}^H\widetilde S^{(N)}.
 \end{split}
 \label{eq:response}
\end{equation}
Replacing $u_{\rm d}^H$ by $u_{\rm d}^{\infty}$ in this overlap gives the ingoing horizon amplitude $B_{{\rm d},H}^{(N)}$.
The integrals are defined in a convergence domain and analytically continued to the required frequencies.

\emph{Recursive IBP at arbitrary order.---}
Inserting the source expansion~\eqref{eq:genericusource} into the overlap $J_N$ in Eq.~\eqref{eq:response} gives integrals containing $q$ lower-order field factors and the homogeneous daughter.

For radiative source factors of orders $\bm p=(p_1,\ldots,p_q)$, write $u_i=u_i^{(p_i)}$ and append the homogeneous daughter as $u_{q+1}=u_{\rm d}^H$.
Only the first $q$ orders sum to $N$; the daughter is an additional Green-function factor.
Define the regulated family
\begin{equation}
 \begin{split}
 I_{\bm p}[\bm d;\bm a;\delta,k]=\int_1^\infty\!\dd r\,
 &e^{-k(r-1)}(r-1)^{\delta-2\ii\omega_{\rm d}-a_h}\\
 &\times r^{-a_r}\prod_\alpha Z_\alpha^{-b_\alpha}
 \prod_{i=1}^{q+1}u_i^{[d_i]}.
 \end{split}
 \label{eq:family}
\end{equation}
Here $\bm d=(d_1,\ldots,d_{q+1})$ records derivative choices and $\bm a=(a_r,a_h,\bm b)$ radial powers. Mode labels are implicit in each family.
The radial weights include $r$, $r-1$, and all distinct factors $Z_\alpha=2\lambda_\alpha r+3$ introduced by the source and its recursive substitutions.
Negative indices represent numerator powers, including the factor $r^5(r-1)$ in Eq.~\eqref{eq:response}.
The analytic regulator $\delta$ shifts endpoint powers; $k$ is an auxiliary parameter used to construct the differential equations and their boundary data.

We suppress $\delta,k$ in the integral labels below.
Integrating the total derivative of the integrand in Eq.~\eqref{eq:family} gives
\begin{equation}
 \begin{split}
 0={}&(\delta-2\ii\omega_{\rm d}-a_h)
 I_{\bm p}[\bm d;a_r,a_h+1,\bm b]\\
 &-kI_{\bm p}[\bm d;\bm a]
 -a_rI_{\bm p}[\bm d;a_r+1,a_h,\bm b]\\
 &-\sum_\alpha2\lambda_\alpha b_\alpha
 I_{\bm p}[\bm d;a_r,a_h,\bm b+\bm e_\alpha]\\
 &+\sum_{i=1}^{q+1}I_{\bm p}[\bm d+\bm e_i;\bm a].
 \end{split}
 \label{eq:ibp}
\end{equation}
A unit vector raises the indicated denominator or derivative index.
Endpoint terms vanish for sufficiently large $\operatorname{Re}\delta$ and $\operatorname{Re}k$; the identity is then analytically continued.

For a radiative lower-order factor, second derivatives are eliminated using
\begin{equation}
 (u_i^{(p_i)})''=-\alpha_i(u_i^{(p_i)})'
 -\beta_i u_i^{(p_i)}+\widetilde S_i^{(p_i)},
 \label{eq:ibpfieldrule}
\end{equation}
where $\alpha_i,\beta_i$ are the coefficients in Eq.~\eqref{eq:unified} for that mode.
The last term is a source of the form in Eq.~\eqref{eq:genericusource}. Insert each of its products in place of the differentiated factor; its rational coefficient shifts the existing radial powers.
The inserted fields have positive orders summing to $p_i$.
Each substitution therefore increases the number of source factors while preserving total order $N$.
There can be at most $N$ source factors, at which point all are first order and their sources vanish.

\emph{Iterative reduction.---}
We write Eq.~\eqref{eq:ibpfieldrule} as a first-order system for $(u_i,u_i')^T$, with homogeneous coefficient matrix
\begin{equation}
 \mathsf C_i(r)=
 \begin{pmatrix}0&1\\-\beta_i(r)&-\alpha_i(r)\end{pmatrix}.
 \label{eq:singlefieldmatrix}
\end{equation}
Leibniz's rule on the $q+1$ pairs $(u_i,u_i')$ gives the homogeneous matrix~\cite{Chen:2023iix,Chen:2024glu}
\begin{equation}
 \begin{split}
 \mathsf C_0+\frac{\mathsf C_r}{r}+\frac{\mathsf C_h}{r-1}
 ={}&\sum_{i=1}^{q+1}\Id_2^{\otimes(i-1)}\otimes\mathsf C_i(r)\\
 &\otimes\Id_2^{\otimes(q+1-i)}.
 \end{split}
 \label{eq:tensorconnection}
\end{equation}
Here $\mathsf C_0$ is the constant term, $\mathsf C_r$ and $\mathsf C_h$ multiply $1/r$ and $1/(r-1)$, and $\Id_2$ is the two-dimensional identity.

Partial fractions separate the radial weights into powers of individual pole factors and polynomial numerators. For example, collect the derivative choices for a fixed product into $\bm I[a_r,a_h]$, with all $b_\alpha=0$.
Including derivatives of the radial weights gives $\mathsf M_0=\mathsf C_0-k\Id$,
$\mathsf M_r(n)=\mathsf C_r-n\Id$, and
$\mathsf M_h=\mathsf C_h+(\delta-2\ii\omega_{\rm d})\Id$, where $\Id$ is the identity of the appropriate dimension.
Equation~\eqref{eq:ibp} gives the double-pole reduction
\begin{equation}
 \begin{split}
 \bm I[2,0]={}&-\mathsf M_r(1)^{-1}\Big[
 (\mathsf M_0-\mathsf M_h)\bm I[1,0]\\
 &+\mathsf M_h\bm I[0,1]+\bm R_1\Big].
 \end{split}
 \label{eq:iterativelowering}
\end{equation}
Here $\bm R_1$ collects source insertions from Eq.~\eqref{eq:ibpfieldrule}. Repeating this step lowers higher pole powers; analogous relations reduce powers of $r-1$, each active $Z_\alpha$, and numerator powers of $r$. At any perturbative order, the single-field equations and the product rule determine the IBP coefficients. The source substitutions terminate at products of first-order fields, completing the reduction rules at that order.

\emph{Counting masters at arbitrary order.---}
Table~\ref{tab:mastercounts} counts a symbolic spanning basis for $N$ distinguishable parents. For generic parameters, the $Z$ factors have distinct roots. Partition the parents into $q\ge2$ groups, each supplying one first-order or intermediate field. Together with the homogeneous daughter, these fields give $2^{q+1}$ derivative products. The closed pole family includes a $Z$ pole for every nonempty union of groups, giving $2^q+1$ simple-pole weights after including $r=0,1$. Each partition therefore contributes $2^{q+1}(2^q+1)$ masters. Summing over all unordered partitions includes the direct interaction ($q=N$) and all cascades.

\begin{table}[htb]
\caption{Generic symbolic spanning counts for distinguishable parents and distinct $Z$ poles, including direct and cascaded contributions.}
\label{tab:mastercounts}
\begin{ruledtabular}
\begin{tabular}{cr}
Order $N$ & Master integrals\\
\hline
$2$ & 40\\
$3$ & 264\\
$4$ & 1,688\\
$5$ & 11,752\\
$6$ & 89,560\\
\end{tabular}
\end{ruledtabular}
\end{table}

\emph{Differential equations.---}
Differentiating Eq.~\eqref{eq:family} with respect to $k$ inserts $-(r-1)$ and lowers the horizon-power index by one:
\begin{equation}
 \partial_k I_{\bm p}[\bm d;a_r,a_h,\bm b]
 =-I_{\bm p}[\bm d;a_r,a_h-1,\bm b].
 \label{eq:kshift}
\end{equation}
Reducing the shifted integrals with the same IBP rules gives a closed system for the master integrals $\bm I$ and expresses the source overlap in this basis:
\begin{equation}
 \frac{\dd\bm I}{\dd k}=\bm\Omega(\delta,k)\bm I,
 \qquad J_N(\delta,k)=\bm C^T(\delta,k)\bm I(\delta,k).
 \label{eq:integralde}
\end{equation}
The matrix $\bm\Omega$ and coefficient vector $\bm C$.

\emph{From integral equations to amplitudes.---}
At large $k$, the factor $e^{-k(r-1)}$ suppresses contributions away from the horizon, so the horizon expansions determine the initial data for the master integrals. After inserting the mode parameters and lower-order boundary amplitudes, we evolve the differential equations~\eqref{eq:integralde} toward $k=0$ and match to a local series.

At fixed noninteger $\delta$, we expand the full source integral $J_N(\delta,k)=\bm C^T\bm I$ near $k=0$ in powers and logarithms of $k$. The divergent terms cancel in the full sum, allowing us to take $k\to0$. The physical overlap is the term independent of $\delta$ in the resulting Laurent expansion about $\delta=0$. We denote extraction of this term by $[\delta^0]$:
\begin{equation}
 J_N=[\delta^0]\left\{\lim_{k\to0}
 \bm C^T(\delta,k)\bm I(\delta,k)\right\}.
 \label{eq:finitepart}
\end{equation}
We first evaluate the limit where the integral converges, then analytically continue the result to a neighborhood of $\delta=0$.

\emph{From the master amplitude to strain.---}
The strain receives a freely propagating contribution proportional to the master amplitude and a local contribution fixed by lower-order fields. In a fixed Bondi frame, the outgoing $r^{-1}$ term defines the strain amplitudes:
\begin{equation}
 \begin{split}
 h_+-\ii h_\times&=\frac1{2r}\sum_{\ell m}A_{\ell m}
 e^{-\ii\omega u_{\rm B}}\,{}_{-2}Y_{\ell m}+\cdots,\\
 A_{\rm d}^{(N)}&=\mathcal N_{\rm d}B_{\rm d}^{(N)}+A_{{\rm d},\rm loc}^{(N)}.
 \end{split}
 \label{eq:strain}
\end{equation}
Here $h_+$ and $h_\times$ are the plus and cross strain polarizations, ${}_{-2}Y_{\ell m}$ are spin-weight $-2$ spherical harmonics, and $u_{\rm B}$ is Bondi retarded time. The standard factor $\mathcal N_{\rm d}$ converts the master amplitude to strain~\cite{Martel:2005ir}, while $A_{{\rm d},\rm loc}^{(N)}$ comes from the local metric terms and lower-order gauge transformations.
To obtain the local contribution, restore $\Psi_{\rm raw}^{(N)}=\Psi^{(N)}+\Delta\Psi^{(N)}$, reconstruct the sourced metric, and transform it to Bondi gauge.

\emph{QNM coupling as an application.---}
Our response kernels become nonlinear QNM coupling coefficients when the parent frequencies are set to $\omega_{\ell n}$~\cite{Leaver:1985ax} and their boundary states are chosen to be ingoing at the horizon and outgoing at infinity.
Here $n=0,1,\ldots$ labels the overtone, with $n=0$ fundamental.
For each parent combination, we quote the strain ratio
\begin{equation}
 \mathcal R=\frac{A_{\rm d}^{(N)}}{A_1\cdots A_N}.
 \label{eq:qnmapplication}
\end{equation}
Here $A_i$ are the first-order parent strain amplitudes, while $A_{\rm d}^{(N)}$ denotes that combination's contribution to the daughter strain.

\begin{figure}[tb]
\centering
\includegraphics[width=\columnwidth]{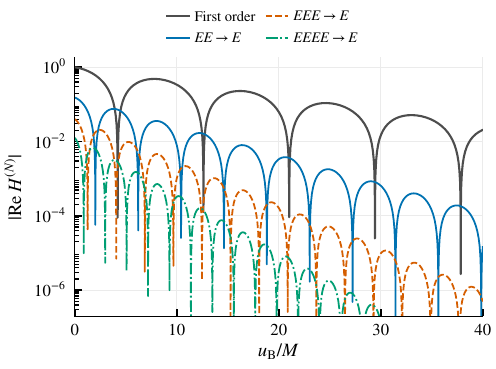}
\caption{All-even waveforms through fourth order. Curves show $|\operatorname{Re}[A_{\rm d}^{(N)}e^{-\ii N\omega_{20}u_{\rm B}}]|$ on a logarithmic scale, with $A_{\rm d}^{(1)}=1$ and $A_{\rm d}^{(N)}=\mathcal R$ for $N=2,3,4$, using the corresponding ratios in Table~\ref{tab:responses}.}
\label{fig:ringdown}
\end{figure}

The quartic source contains $1+1+1+1$, $2+1+1$, $2+2$, and $3+1$ contributions; the $2+2$ term couples two driven second-order fields.
Figure~\ref{fig:ringdown} shows the all-even sequence through fourth order.

\begin{table}[tb]
\caption{QNM strain ratios $\mathcal R$, defined in Eq.~\eqref{eq:qnmapplication}, for all allowed parity channels of $N=2,3,4$ fundamental $(\ell,m,n)=(2,2,0)$ parents. The daughter has $(\ell,m)=(2N,2N)$ and frequency $N\omega_{20}$. All distinct parent orderings are included.}
\label{tab:responses}
\begin{ruledtabular}
\begin{tabular}{lrr}
Channel & $\operatorname{Re}\mathcal R$ & $\operatorname{Im}\mathcal R$\\
\hline
$EE\to E$   & $0.153122626964$ & $-0.010443911003$\\
$EO\to O$   & $0.273189109379$ & $-0.022877632660$\\
$OO\to E$   & $0.120066482415$ & $-0.012433721657$\\
$EEE\to E$  & $0.040344314019$ & $-0.006534012981$\\
$EEO\to O$  & $0.099940091380$ & $-0.017420086727$\\
$EOO\to E$  & $0.083342996792$ & $-0.014816174448$\\
$OOO\to O$  & $0.023747219431$ & $-0.003930100702$\\
$EEEE\to E$ & $0.012648677421$ & $-0.003424514975$\\
$EEEO\to O$ & $0.039309603747$ & $-0.010902362667$\\
$EEOO\to E$ & $0.047084282959$ & $-0.012800274842$\\
$EOOO\to O$ & $0.025809444134$ & $-0.006673989483$\\
$OOOO\to E$ & $0.005386087502$ & $-0.001351562333$\\
\end{tabular}
\end{ruledtabular}
\end{table}

An extension of this framework to Kerr black holes at arbitrary perturbative order will be presented in forthcoming work~\cite{Chen:KerrResponse}.

\begin{acknowledgments}
Y.-C.~X. thanks Bangalore Sathyaprakash for helpful discussions. J.~C. is supported by the National Natural Science Foundation of China (NSFC) under Grant No.~12505094 and by the Science Foundation of China University of Petroleum, Beijing (No.~2462025YJRC019). Y.-C.~X. acknowledges support from the U.S. National Science Foundation (NSF) under Grant No.~PHY-2608011.

Y.-C.~X. used ACES at Texas A\&M University, supported by NSF Award No.~2112356, through ACCESS allocation PHY260122. The ACCESS program is supported by NSF Grant Nos.~2138259, 2138286, 2138307, 2137603, and 2138296~\cite{Boerner:2023acc}. Y.-C.~X. also used the Roar supercomputer (RRID:SCR\_026424), provided by the Institute for Computational and Data Sciences at The Pennsylvania State University.

The authors developed the scientific ideas and methodology. GPT~5.5, GPT~5.6~sol, and GPT~6~astra assisted with part of the code implementation and refactoring. The authors are responsible for the scientific content, calculations, and conclusions.
\end{acknowledgments}

\emph{Data and code availability.---}
The symbolic ODE system for master integrals, together with the code to evaluate it, will be released at \url{https://github.com/YucunXie/Nonlinear_Response_RWZ}.

\bibliography{refs}

\onecolumngrid
\newpage
\twocolumngrid
\appendix
\setcounter{equation}{0}
\renewcommand{\theequation}{S\arabic{equation}}
\section*{End Matter}
\label{sec:supplemental}

\emph{1. Constructing the source completion.---}
To determine the completion, we match $\mathcal L_{\rm d}\Delta\Psi^{(N)}$ to the leading boundary terms in $S_{\rm raw}^{(N)}$. We build $\Delta\Psi^{(N)}$ from known lower-order fields, with trial terms of the form
\begin{equation}
 \Delta\Psi_{\rm trial}^{(N)}
 =c\,w(r)\prod_i\bigl(\Psi_i^{(p_i)}\bigr)^{[d_i]}.
 \label{eq:completionansatz}
\end{equation}
Here $c$ is an unknown coefficient independent of $r$, and $w(r)$ is a rational radial weight. The field orders satisfy $1\le p_i<N$ and $\sum_i p_i=N$, while $d_i=0,1$ selects the field or its first derivative.

The coefficients of different trial products are coupled because the lower-order fields can themselves have sources. At third order, this links the $2+1$ and $1+1+1$ terms. Eliminating second derivatives with the sourced RWZ equations gives
\begin{equation}
 \begin{split}
 \mathcal L_{\rm d}\bigl(\Psi_1^{(2)}\Psi_2^{(1)}\bigr)
 ={}&2(D_*\Psi_1^{(2)})(D_*\Psi_2^{(1)})\\
 &+(V_1+V_2-V_{\rm d}+2\omega_1\omega_2)
       \Psi_1^{(2)}\Psi_2^{(1)}\\
 &+S_1^{(2)}\Psi_2^{(1)}.
 \end{split}
 \label{eq:completionproduct}
\end{equation}
Here $\Psi_1^{(2)}$ and $\Psi_2^{(1)}$ have frequencies $\omega_1,\omega_2$ and RWZ potentials $V_1,V_2$. The daughter has $\omega_{\rm d}=\omega_1+\omega_2$ and potential $V_{\rm d}$. The last term contains the quadratic source $S_1^{(2)}$ of $\Psi_1^{(2)}$, and therefore expands into products of three first-order fields.

After eliminating second derivatives, we expand $\mathcal L_{\rm d}\Delta\Psi^{(N)}$ and $S_{\rm raw}^{(N)}$ in powers of $r-1$ near the horizon and $1/r$ near infinity. Matching their leading coefficients for each independent field product gives a linear system for $c$. For incoming or mixed first-order parents, both wave components and their products enter this matching and the metric reconstruction.

For example, take $S_{\rm raw}^{(N)}=e^{\ii\omega_{\rm d}r_*}[S_\infty+O(r^{-1})]$ at infinity, with $\omega_{\rm d}\ne0$. Here $S_\infty$ is the leading outgoing source coefficient. By frequency conservation, the outgoing field products in Eq.~\eqref{eq:completionansatz} have the same phase. Weights with $w(r)\sim r$ therefore supply a completion term $c_\infty r e^{\ii\omega_{\rm d}r_*}$, where $c_\infty$ combines the trial coefficients with the known field amplitudes. Applying the daughter operator gives
\begin{equation}
 \mathcal L_{\rm d}\bigl(r e^{\ii\omega_{\rm d}r_*}\bigr)
 =e^{\ii\omega_{\rm d}r_*}
   \bigl[2\ii\omega_{\rm d}+O(r^{-1})\bigr],
 \label{eq:completionbalance}
\end{equation}
so the corrected source $S^{(N)}=S_{\rm raw}^{(N)}-\mathcal L_{\rm d}\Delta\Psi^{(N)}$ contains $(S_\infty-2\ii\omega_{\rm d}c_\infty)e^{\ii\omega_{\rm d}r_*}$ at leading order. Cancelling this term fixes
\[
 S_\infty-2\ii\omega_{\rm d}c_\infty=0,
 \qquad c_\infty=\frac{S_\infty}{2\ii\omega_{\rm d}}.
\]
We add no independent homogeneous solution to $\Delta\Psi^{(N)}$. For our QNM examples, the completed source has the asymptotic form $S^{(N)}=e^{\ii\omega_{\rm d}r_*}O(r^{-2})$ at infinity.

\emph{2. Radial conversion and boundary amplitudes.---}
With the completion fixed, we transform the source to the common variable $u^{(N)}$. The Darboux function and normalization in Eq.~\eqref{eq:ustrip} are
\begin{equation}
 \begin{gathered}
 \mathcal D(r)=\frac{2\lambda(\lambda+1)}3+\frac{3(r-1)}{r^2Z},\\
 \kappa=\left[\frac{2\lambda(\lambda+1)}3\right]^2+\omega^2.
 \end{gathered}
 \label{eq:darboux}
\end{equation}
The transformed source is
\begin{equation}
 \widetilde S^{(N)}=\frac{r^{-3}(r-1)^{\ii\omega}}{f^2}
 \begin{cases}
 S^{(N)},&P=O,\\
 (-D_*+\mathcal D)S^{(N)}/\kappa,&P=E
 \end{cases}.
 \label{eq:usource}
\end{equation}
Initial data at large $k$ come from the horizon expansions of the master integrands. We form these from the lower-order fields, the homogeneous daughter, and the radial weights. The lower-order fields contribute both their known horizon amplitudes and the terms fixed by their sources. Integrating each term against $e^{-k(r-1)}$ gives the large-$k$ expansion (Watson's lemma~\cite{NIST:2026dlm}):
\begin{equation}
 \int_1^\infty\dd r\,e^{-k(r-1)}(r-1)^{a+\delta}
 =\frac{\Gamma(a+\delta+1)}{k^{a+\delta+1}}.
 \label{eq:watsonseed}
\end{equation}
Here $a$ is the power of $r-1$ before adding the regulator. Logarithmic factors are included by differentiating this identity with respect to $a$.

We evolve the master integrals toward $k=0$. For our QNM examples, the contour passes below poles on the positive real $k$ axis. We assemble $\bm C^T\bm I$ at fixed $\delta$ and take its ordinary $k\to0$ limit. In these examples, the part of the large-$r$ integrand whose wave phases cancel behaves as $r^{-2+\delta}$, so it is integrable at infinity near $\delta=0$. After the divergent terms cancel in the full sum, its constant term in $k$ gives the limit.

Only then do we extract $[\delta^0]$, as in Eq.~\eqref{eq:finitepart}. The coefficient and master Laurent series must be multiplied before this extraction: a $1/\delta$ coefficient times a term proportional to $\delta$ in a master integral contributes to the constant term.

On a frequency branch cut, we evaluate the limits from its two sides separately. At a simple daughter QNM pole, the Wronskian vanishes. We expand the full response numerator and Wronskian about that frequency to obtain the pole coefficient and regular term. Together with the homogeneous solution fixed by the physical initial conditions, these determine the resonant contribution to the time-domain response.

\emph{3. Reconstructing the waveform.---}
The Green-function integral gives $B_{\rm d}^{(N)}$. To obtain the strain, we also need the metric terms fixed by lower-order fields. We first recover the boundary expansion of the completed master field $\Psi^{(N)}$ from $u^{(N)}$. In odd parity, $\Psi^{(N)}=r^3(r-1)^{-\ii\omega}u^{(N)}$. For even parity,
\begin{equation}
 \Psi^{(N)}=(D_*+\mathcal D)
 \bigl[r^3(r-1)^{-\ii\omega}u^{(N)}\bigr]+\frac{S^{(N)}}{\kappa}.
 \label{eq:sourcedmap}
\end{equation}
At infinity, $\Psi_{\rm d}^{\infty}$ is the unit-amplitude outgoing homogeneous solution. We assign the free-wave term to $B_{\rm d}^{(N)}\Psi_{\rm d}^{\infty}$ and collect the terms determined by $S^{(N)}$ in $\Psi_{\rm source}^{(N)}$. Restoring the completion gives
\begin{equation}
 \Psi_{\rm raw}^{(N)}=B_{\rm d}^{(N)}\Psi_{\rm d}^{\infty}
                  +\Psi_{\rm source}^{(N)}+\Delta\Psi^{(N)}.
 \label{eq:rawreadout}
\end{equation}
In even parity, $\Psi_{\rm source}^{(N)}$ includes the local $S^{(N)}/\kappa$ term in Eq.~\eqref{eq:sourcedmap}.

Powers of $r$ in the metric reconstruction allow subleading master-field terms to contribute to the leading strain. For our QNM examples, write
\begin{equation}
 \begin{split}
 S^{(N)}&=e^{\ii\omega_{\rm d}r_*}
       \left(\frac{s_2}{r^2}+\frac{s_3}{r^3}+\cdots\right),\\
 \Psi_{\rm source}^{(N)}&=e^{\ii\omega_{\rm d}r_*}
       \left(\frac{\psi_1}{r}+\frac{\psi_2}{r^2}+\cdots\right).
 \end{split}
 \label{eq:sourceseries}
\end{equation}
The lower-order boundary expansions supply $s_j$. Matching powers of $1/r$ in $\mathcal L_{\rm d}\Psi_{\rm source}^{(N)}=S^{(N)}$ then determines $\psi_j$ successively. The first relation is
\begin{equation}
 \psi_1=-\frac{s_2}{2\ii\omega_{\rm d}}.
 \label{eq:firstsourcecoefficient}
\end{equation}

We reconstruct the radiative metric $h^{(N)}$ for $\ell\ge2$ from $\Psi_{\rm raw}^{(N)}$ and $\mathcal T^{(N)}$~\cite{Spiers:2023mor}. The formulas contain terms linear in the raw master and explicit terms from the projected Einstein source. We insert the boundary series into both sets of terms, retaining enough radial orders to determine the coefficient of $r$ in the angular metric.

The strain is determined by the radiative angular metric in Bondi gauge. We transform to this gauge in retarded coordinates $(t-r_*,r,\theta,\varphi)$. At order $N$, we solve for a new gauge vector $\xi^{(N)}$. The terms involving lower-order fields and gauge vectors are already known; we collect them in $\Delta h_{\rm lower}^{(N)}$. The transformed metric is~\cite{Bruni:1996im}
\begin{equation}
 h_{\rm B}^{(N)}=h^{(N)}+\pounds_{\xi^{(N)}}g_0
                 +\Delta h_{\rm lower}^{(N)}.
 \label{eq:bondirecursion}
\end{equation}
Here $\pounds_\xi$ denotes a Lie derivative. At second order, the known contribution is
\begin{equation}
 \Delta h_{\rm lower}^{(2)}
 =\pounds_{\xi^{(1)}}h^{(1)}
   +\frac12\pounds_{\xi^{(1)}}^2g_0.
 \label{eq:lowergaugeexample}
\end{equation}

Let $\gamma_{AB}$ be the unit-sphere metric, with $A,B$ angular indices. We determine $\xi^{(N)}$ by imposing the Bondi conditions on the full transformed metric $g_{\rm B}$~\cite{Madler:2016xju}:
\begin{equation}
 g_{{\rm B},rr}=g_{{\rm B},rA}=0,\qquad
 \det(g_{{\rm B},AB})=r^4\det(\gamma_{AB}).
 \label{eq:bondiconditions}
\end{equation}
At order $N$, these are linear equations for $\xi^{(N)}$ because the new vector acts only on $g_0$. The determinant equation also contains known products of lower-order metric terms. We fix the integration constants so that $g_{{\rm B},u_{\rm B}r}=-1+O(r^{-2})$ and the leading angular metric is $r^2\gamma_{AB}$, using the same residual Bondi frame at every order.

Dividing the following angular metric combination by $2r^2$ gives the complex strain:
\begin{equation}
 \begin{split}
 &h_{{\rm B},\theta\theta}^{(N)}
 -\frac{h_{{\rm B},\varphi\varphi}^{(N)}}{\sin^2\theta}
 -\frac{2\ii h_{{\rm B},\theta\varphi}^{(N)}}{\sin\theta}\\
 &\qquad=r\sum_{\ell m}A_{\ell m}^{(N)}
       e^{-\ii\omega u_{\rm B}}\,{}_{-2}Y_{\ell m}+O(1).
 \end{split}
 \label{eq:angularreadout}
\end{equation}
Thus the outgoing coefficient of $r$, with no $\log r$ factor, gives the strain amplitudes in Eq.~\eqref{eq:strain} after harmonic projection. Leading logarithmic strain coefficients are time-independent, so they vanish for $\omega_{\rm d}\ne0$~\cite{Chrusciel:1995polyhom}. With orthonormal spin-weighted harmonics and our odd-parity convention, the conversion factors are
\begin{equation}
 \mathcal N_\ell^{E}=\sqrt{\frac{(\ell+2)!}{(\ell-2)!}},\qquad
 \mathcal N_\ell^{O}=-\ii\mathcal N_\ell^{E}.
 \label{eq:linearstrain}
\end{equation}
The reconstructed strain contains a term proportional to $B_{\rm d}^{(N)}$ and a term independent of it, $A_{{\rm d},\rm loc}^{(N)}$. We compute $A_{{\rm d},\rm loc}^{(N)}$ by repeating the reconstruction and Bondi transformation with $B_{\rm d}^{(N)}=0$, while holding all lower-order amplitudes fixed. This includes contributions from $\Psi_{\rm source}^{(N)}$, $\Delta\Psi^{(N)}$, the explicit metric-source terms, and $\Delta h_{\rm lower}^{(N)}$. A change of completion shifts the propagated and local contributions oppositely, leaving the strain invariant.

\emph{4. Monopole and dipole sectors.---}
The nonradiative $\ell=0,1$ perturbations are determined from the gauge-fixed Einstein constraints~\cite{Spiers:2023mor}. Their homogeneous parts describe a mass shift in the monopole, an angular-momentum shift in the odd dipole, and a change of origin or rest frame in the even dipole. We fix these homogeneous components, retain the contributions driven by lower-order fields, and include the resulting perturbations in subsequent nonlinear sources and waveform reconstruction.

\end{document}